\documentclass[pdflatex,sn-mathphys-num]{sn-jnl}% Math and Physical Sciences Numbered Reference Style
\usepackage{graphicx}%
\usepackage{multirow}%
\usepackage{amsmath,amssymb,amsfonts}%
\usepackage{amsthm}%
\usepackage{mathrsfs}%
\usepackage[title]{appendix}%
\usepackage{xcolor}%
\usepackage{textcomp}%
\usepackage{manyfoot}%
\usepackage{booktabs}%
\usepackage{algorithm}%
\usepackage{algorithmicx}%
\usepackage{algpseudocode}%
\usepackage{listings}%
\usepackage{placeins}%
\theoremstyle{thmstyleone}%
\theoremstyle{thmstyletwo}%

\theoremstyle{thmstylethree}%

\begin{document}

\title[Composable multi-satellite precipitation estimation]{Composable multi-satellite precipitation estimation for evolving observing systems}


\author[1,2,3]{\fnm{Yunfan} \sur{Yang}}\email{yangyunfan22@mails.ucas.ac.cn}
\equalcont{These authors contributed equally to this work.}

\author[4]{\fnm{Haofei} \sur{Sun}}\email{sunhf@typhoon.org.cn}
\equalcont{These authors contributed equally to this work.}

\author[3]{\fnm{Xiuyu} \sur{Sun}}\email{xiuyu.sxy@gmail.com}

\author*[5,6]{\fnm{Wei} \sur{Han}}\email{hanwei@cma.gov.cn}

\author*[3]{\fnm{Xiaoze} \sur{Xu}}\email{xuxiaoze@sais.org.cn}

\author[7]{\fnm{Xingtao} \sur{Song}}\email{202511050004@nuist.edu.cn}

\author[8]{\fnm{Jun} \sur{Li}}\email{junli@cma.gov.cn}

\author[1,2]{\fnm{Zhiqiu} \sur{Gao}}\email{zgao@mail.iap.ac.cn}

\author[4]{\fnm{Wei} \sur{Huang}}\email{huangw@typhoon.org.cn}

\affil*[1]{\orgdiv{State Key Laboratory of Atmospheric Boundary Layer Physics and Atmospheric Chemistry (LAPC)}, \orgname{Institute of Atmospheric Physics, Chinese Academy of Sciences}, \orgaddress{\city{Beijing}, \country{China}}}

\affil[2]{\orgdiv{College of Earth and Planetary Sciences}, \orgname{University of Chinese Academy of Sciences}, \orgaddress{\city{Beijing}, \country{China}}}

\affil[3]{\orgname{Shanghai Academy of Artificial Intelligence for Science (SAIS)}, \orgaddress{\city{Shanghai}, \country{China}}}

\affil[4]{\orgdiv{Key Laboratory of Numerical Modeling for Tropical Cyclone of the China Meteorological Administration}, \orgname{Shanghai Typhoon Institute}, \orgaddress{\city{Shanghai}, \country{China}}}

\affil[5]{\orgdiv{State Key Laboratory of Severe Weather}, \orgname{Chinese Academy of Meteorological Sciences}, \orgaddress{\city{Beijing}, \country{China}}}

\affil[6]{\orgname{CMA Earth System Modeling and Prediction Centre (CEMC)}, \orgaddress{\city{Beijing}, \country{China}}}

\affil[7]{\orgdiv{School of Atmospheric Physics}, \orgname{Nanjing University of Information Science and Technology}, \orgaddress{\city{Nanjing}, \country{China}}}

\affil[8]{\orgdiv{National Satellite Meteorological Center}, \orgname{China Meteorological Administration}, \orgaddress{\city{Beijing}, \country{China}}}


\abstract{Rapid and spatially continuous precipitation monitoring is critical for flood, landslide, and other hydrometeorological hazard warnings, particularly in regions where rain-gauge and weather-radar networks are sparse. The coordinated use of heterogeneous satellite observations, including geostationary infrared, passive microwave, and spaceborne radar measurements, is therefore a key pathway toward more accurate and spatially refined precipitation monitoring. Recent deep-learning methods have substantially improved multi-source satellite precipitation estimation, but most remain tied to predefined combinations of satellite inputs. As satellite observing systems evolve, incorporating new instruments often requires substantial model retraining and maintenance. We propose PRISMA, a generative framework for precipitation retrieval from multi-source observations. The framework separates the training of the precipitation prior from sensor-specific observational constraints, enabling sensor branches to be flexibly composed or extended without retraining the precipitation generative backbone. We successively integrate FY-4B/AGRI, GPM/GMI, F16--F18 SSMIS, and GPM/DPR-Ka observations within the PRISMA framework, achieving consistent improvements in precipitation-estimation accuracy. Matched-footprint experiments further confirm the effective use of complementary information from coincident sensors. Independent station validation shows that PRISMA outperforms IMERG Final in both CRPS and RMSE while providing positive fair Brier skill across all precipitation thresholds. PRISMA enables flexible composition of heterogeneous satellite observations and rapid generation of accurate ensemble precipitation estimates, strengthening satellite-based precipitation monitoring.}

\keywords{Satellite precipitation estimation, Multi-satellite observations, Generative modeling, Observation conditioning, Probabilistic retrieval, Evolving observing systems}



\maketitle

\section{Introduction}\label{introduction}

Timely and spatially continuous precipitation information is fundamental to warning systems for floods, landslides and other hydrometeorological hazards, as well as to water-resource and agricultural management \cite{wmo2021atlas,ipcc2021ar6,dai2025precipdiff}. Yet precipitation remains difficult to observe at the coverage and frequency required for rapid decision-making, particularly in regions where rain-gauge and weather-radar networks are sparse or unevenly distributed \cite{brempong2025oya,wmo2021water,yuval2026neural}. A recent global assessment estimated that only 13.4\% of the land surface meets World Meteorological Organization requirements for annual precipitation monitoring, with particularly large observational gaps across Africa, Central Asia and South America \cite{su2026gauge}. Satellite observations are therefore indispensable for extending quantitative precipitation monitoring beyond well-instrumented regions. For hazard-related applications, effective monitoring also requires timely and accurate estimates of precipitation intensity and spatial distribution \cite{zhang2023nowcastnet}.

A single satellite cannot fully satisfy these requirements. Geostationary infrared imagers provide frequent and spatially continuous observations, but mainly characterize cloud-top thermal properties and therefore provide only indirect information on whether precipitation occurs and how intense it is. Passive microwave radiometers and spaceborne precipitation radars are more directly sensitive to hydrometeors, but their observations are confined to intermittent orbital swaths \cite{arkin1987relationship,kummerow1998trmm,hou2014gpm}. Multi-satellite precipitation estimation must therefore exploit measurements with complementary physical sensitivity while accommodating large differences in spatial coverage, sampling time and instantaneous availability. These differences become increasingly important as satellite constellations expand because the observations available for any retrieval can vary across both space and time. The central challenge is therefore not simply to fuse more channels, but to infer precipitation consistently from whatever combination of observations is actually available.

Satellite precipitation retrieval has long been treated as a probabilistic inverse problem in which prior knowledge of plausible atmospheric states is combined with satellite measurements to constrain the range of possible solutions. This principle underlies Bayesian passive-microwave retrieval algorithms such as GPROF, which uses an a priori database of precipitation profiles and associated brightness temperatures to constrain retrievals from observed microwave radiances \cite{gprof_atbd,huffman2020imerg}. Much of modern supervised deep learning has replaced this explicit separation between prior information and observational constraints with an end-to-end conditional mapping from satellite measurements to retrieval targets \cite{guilloteau2024generative,li2025latent,tu2025mods}. Although these models can capture highly nonlinear relationships and represent retrieval uncertainty, the learned target distribution remains tied to the observation configuration used during training. This limits their extensibility to new or changing sensor combinations, which generally require modification or retraining of the conditional model. Such dependence is particularly restrictive for multi-satellite precipitation estimation, where observation availability varies continuously across space and time as orbital swaths intersect and satellite constellations evolve.

Generative models provide a data-driven way to recover the separation between a target prior and observational constraints. In score-based inverse problems, an unconditional score model can first be learned from the target distribution and then combined at inference time with gradients derived from a known observation operator and an observation-error or likelihood model, allowing new measurements to constrain the sampling process without retraining the prior \cite{song2022solving,chung2023diffusion}. This formulation closely parallels Bayesian retrieval and data assimilation, in which a background or prior distribution is updated by observations through an observation operator and an associated error model. Its direct application to multi-satellite precipitation retrieval is challenging. Infrared radiances, microwave brightness temperatures and radar reflectivities are linked to precipitation through distinct and highly nonlinear radiative-transfer and microphysical processes, making accurate observation operators and likelihood models difficult to construct for every sensor. This difficulty is well recognized in satellite data assimilation, where cloud- and precipitation-affected observations have long posed challenges because of strongly nonlinear forward operators, non-Gaussian and state-dependent errors, and uncertainties in moist-physics and radiative-transfer modeling \cite{errico2007issues,bauer2011satellite,geer2018allsky}. Gradient-based generative inference introduces an additional requirement that the relevant observation operators be differentiable. Building and maintaining such a set of sensor-specific operators for heterogeneous satellite observations would therefore introduce substantial physical-modeling and engineering complexity.

Here we introduce \textbf{PRISMA}, a composable generative framework for multi-satellite precipitation estimation (\textbf{Fig. 1}). Inspired by the conditional-control strategy of ControlNet \cite{zhang2023controlnet}, PRISMA separates a reusable precipitation prior from sensor-specific observational constraints. To enable efficient learning over large-domain precipitation fields and heterogeneous satellite observations, precipitation and each observation source are first encoded into compact latent representations using a precipitation tokenizer and sensor-specific observation tokenizers. PRISMA then learns a precipitation generative backbone in latent space and freezes it, while independent conditioning branches learn how individual observation types constrain the shared precipitation distribution. At inference time, the branches corresponding to the observations that are actually available can be activated jointly, with spatial weight maps assigning different weights to the sensor-specific constraints before their combination within the frozen generative backbone. Conceptually, the inference transforms the learned precipitation distribution \(p(y)\) into an observation-conditioned distribution \(p(y\mid\mathcal{O})\), where \(y\) denotes the precipitation field and \(\mathcal{O}\) denotes the available observations.

In this study, we integrate four heterogeneous satellite observing components within PRISMA: FY-4B/AGRI geostationary infrared imagery, GPM/GMI passive microwave radiances, a merged F16--F18 SSMIS microwave constellation, and GPM/DPR-Ka active radar observations. Across July and August 2025, successive sensor integration consistently improves both deterministic and probabilistic precipitation estimation, with full-domain gains increasing with instantaneous microwave coverage. Matched-footprint experiments further show that coincident sensors provide complementary information beyond expanded spatial coverage. Independent validation against China's national rain-gauge network confirms the incremental benefit of multi-source integration, with the four-source configuration outperforming IMERG Final in both RMSE and CRPS while retaining positive fair Brier skill across the evaluated precipitation thresholds. Overall, these results demonstrate that PRISMA can incorporate heterogeneous satellite observations incrementally while preserving a shared precipitation backbone, providing a flexible framework for multi-satellite precipitation monitoring.

\begin{figure}
\centering
\includegraphics[width=0.95\linewidth,height=\textheight,keepaspectratio]{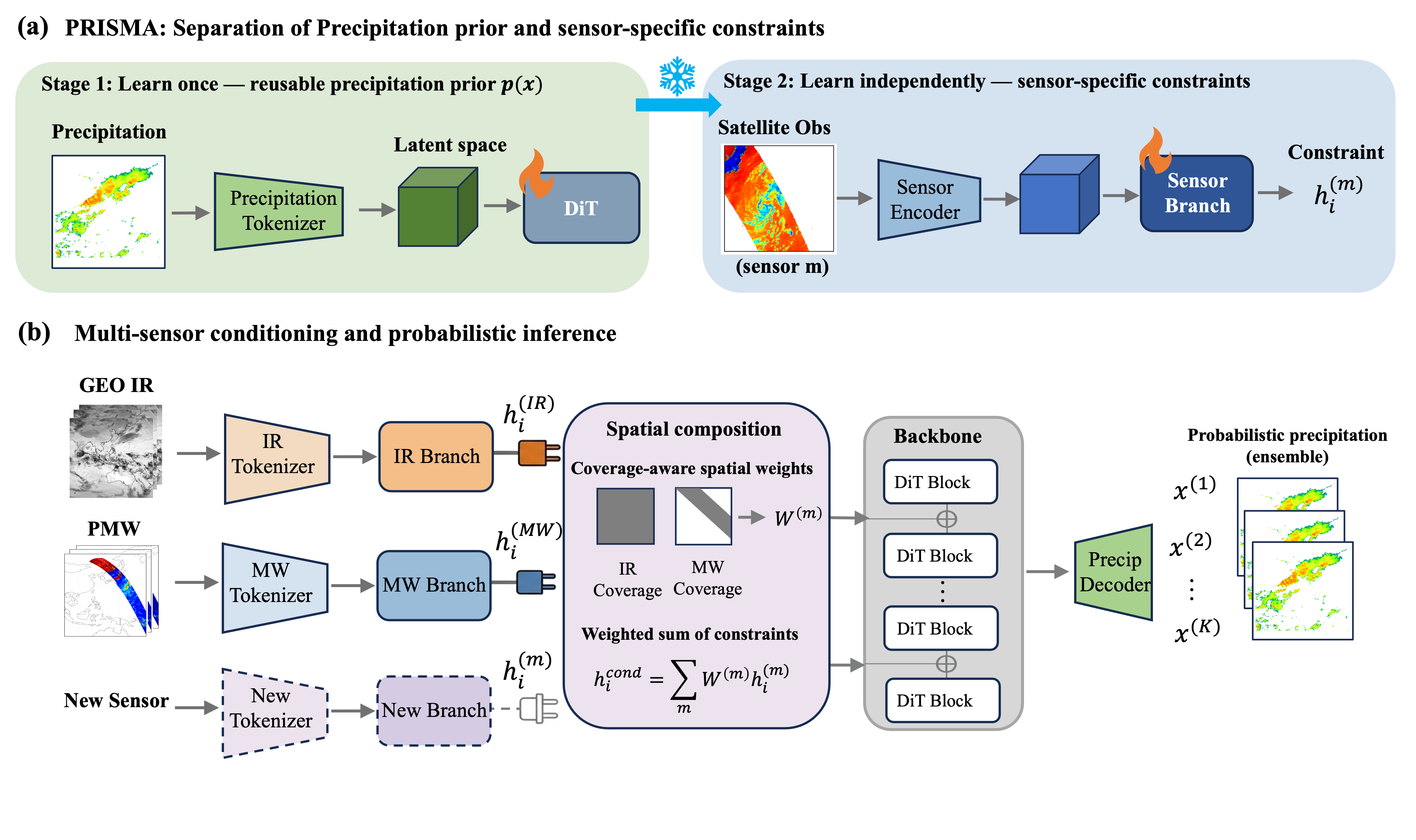}
\caption{Overview of PRISMA. a, PRISMA decouples a reusable precipitation prior from sensor-specific observational constraints. The latent generative backbone is first trained on precipitation fields and then frozen, while independent conditional branches learn how individual satellite observations constrain the shared prior. b, At inference, constraints from the available sensor branches are spatially composed according to observation coverage and injected layer-wise into the frozen backbone. Additional sensors can be incorporated through new independently trained branches without retraining the precipitation prior, and repeated conditional sampling yields an ensemble of spatially continuous precipitation estimates.}\label{fig:framework}
\end{figure}

\section{Results}\label{results}

\subsection{Evaluation of latent representations and the precipitation prior}\label{evaluation-of-latent-representations-and-the-precipitation-prior}

The quality of the latent representation directly affects the ability of the generative backbone to learn precipitation structure and observational constraints. While tokenizers pre-trained on large-scale natural imagery provide a strong initialization with baseline encoding capabilities, their general-purpose representations must be adapted to capture the distinct physical properties of meteorological fields. By leveraging these visual tokenizers as a robust starting point for domain-specific fine-tuning, we examined whether the adapted latent representations could successfully preserve the critical structures of both the precipitation target and the multi-sensor inputs.

A dedicated tokenizer was inherently required for the target variable, as precipitation fields are sparse, strongly intermittent, and dominated by a long-tailed intensity distribution that fundamentally differs from natural imagery. We therefore first examined whether a vision-pretrained tokenizer could faithfully represent precipitation fields, and to what extent domain adaptation improved the latent representation. Across a 1,200 \(\times\) 1,200-grid domain during July and August 2025, domain fine-tuning consistently improved reconstruction accuracy (\textbf{Table S1}). Domain fine-tuning reduced the MAE by 65.6\%, from 0.064 to 0.022 mm h\(^{-1}\), and reduced the RMSE by 48.0\%, from 0.294 to 0.153 mm h\(^{-1}\). Improvements also extended across all evaluated precipitation thresholds. CSI increased from 0.654 to 0.898 at 0.1 mm h\(^{-1}\) and from 0.803 to 0.900 at 1 mm h\(^{-1}\). At higher thresholds, CSI increased from 0.682 to 0.831 at 5 mm h\(^{-1}\) and from 0.600 to 0.774 at 10 mm h\(^{-1}\).

The representative cases in \textbf{Fig. 2a--c} illustrate the origin of these improvements. Domain adaptation reduced the reconstruction RMSE from 0.327 to 0.173 mm h\(^{-1}\) and increased CSI at the 1 mm h\(^{-1}\) threshold from 0.810 to 0.903 (\textbf{Fig. 2a--c}). Whereas the vision-pretrained tokenizer produced spurious weak-precipitation signals over otherwise dry regions, the adapted tokenizer largely suppressed these artefacts while retaining the spatial organization and intensity hierarchy of the precipitation systems. This enhancement is critical for the subsequent generative model; systematic reconstruction errors at this stage could otherwise become embedded in the learned precipitation distribution and propagate into sensor-conditioned retrievals.

Domain adaptation similarly improved the representation of heterogeneous satellite observations. For the three AGRI channels directly matched to the vision-pretrained tokenizer, fine-tuning reduced the mean RMSE by 11.3\%, from 2.405 to 2.134 K, reduced the mean MAE by 13.3\%, from 1.623 to 1.407 K, reduced the mean absolute MBE by 73.2\%, from 0.779 to 0.209 K, and increased the mean SSIM from 0.954 to 0.964 (\textbf{Table S2a}). For GMI, the comparison was restricted to three representative high-frequency channels to match the three-channel input configuration of the vision-pretrained tokenizer. Domain adaptation produced particularly large gains for these channels: RMSE decreased by 84.6--90.3\% and MAE by 79.2--84.6\%, absolute MBE decreased from 2.34--3.94 K to 0.14--0.26 K, and SSIM increased from 0.934--0.954 to 0.962--0.975 (\textbf{Table S3a}). Reconstruction performance across all 13 GMI channels is reported in \textbf{Table S3b}. The adapted SSMIS and DPR-Ka tokenizers likewise reduced reconstruction errors relative to their vision-pretrained initializations, with complete statistics reported in \textbf{Tables S4--S5}.

These dataset-wide improvements were also apparent in individual satellite observations (\textbf{Fig. 2d--i}). For the AGRI 10.8-\(\mu\)m channel, domain adaptation reduced the RMSE from 3.390 to 3.154 K and increased SSIM from 0.927 to 0.939 (\textbf{Fig. 2d--f}). The effect was considerably stronger for high-frequency microwave observations. For the GMI 166.0-V channel, fine-tuning reduced RMSE from 40.604 to 5.084 K and increased SSIM from 0.910 to 0.944 (\textbf{Fig. 2g--i}). Although the vision-pretrained tokenizer retained some capacity to encode broad image structure, it substantially distorted the narrow-swath microwave observations and their brightness-temperature variability. Domain adaptation recovered coherent brightness-temperature gradients and scattering-related spatial structures, demonstrating that modality-specific adaptation is essential for constructing a latent representation that preserves the distinct information content of heterogeneous satellite sensors.

\begin{figure}
\centering
\includegraphics[width=0.95\linewidth,height=\textheight,keepaspectratio]{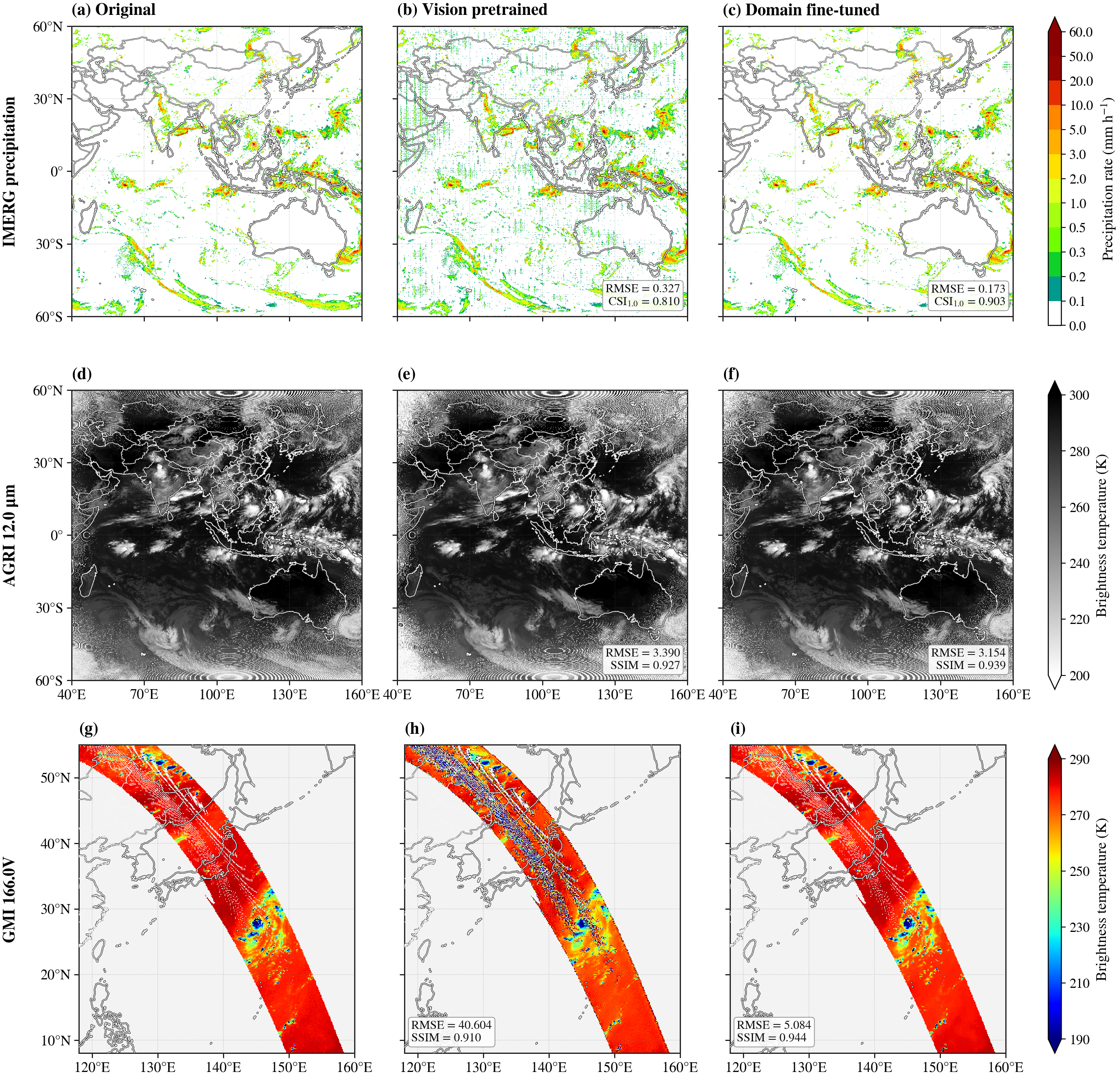}
\caption{Tokenizer reconstruction fidelity across precipitation, infrared, and microwave observations. (a)-(c) show the original, vision-pretrained reconstruction, and domain-fine-tuned reconstruction of the IMERG precipitation-rate field, respectively. (d)-(f) present the corresponding results for the AGRI 10.8\(\mu\)m brightness-temperature field, while (g)-(i) show those for the GMI 166.0-V channel.}\label{fig:tokenizer}
\end{figure}

Having established that the latent representation preserves both the observational signals and the intermittent structure of precipitation, we next evaluated whether the unconditional rectified-flow backbone could learn a robust precipitation prior. We compared 500 unconditionally generated fields with 500 randomly sampled fields from the 2022--2024 IMERG Final archive, utilizing spatial climatology alongside sample-level distributions of the field mean and variance as complementary diagnostics (\textbf{Fig. 3}). The generated prior successfully reproduced the principal climatological gradients of the reference---including the Intertropical Convergence Zone, the Western Pacific Warm Pool, and the South and Southeast Asian monsoon maxima---in both spatial mean (\textbf{Fig. 3a,b}) and spatial variance (\textbf{Fig. 3c,d}). Notably, the generated fields retained regions of intense convective variability rather than over-smoothing them into a uniform background. At the sample level, the generated distributions of the field mean and variance were statistically indistinguishable from the reference data at conventional significance levels (Kolmogorov--Smirnov test, p=0.111 and p=0.095, respectively; \textbf{Fig. 3e,f}), and the variance histograms exhibited a high Pearson correlation of 0.907.

This distributional agreement is primarily attributable to the two-phase noise-scale curriculum rather than merely extending the training duration. The model checkpoint trained solely with a high noise-scale shift (s=5 for 50,000 iterations) produced field-mean and variance distributions significantly narrower than the reference (KS p=0.002 and p=0.003, respectively), indicating insufficient sample diversity. The subsequent 20,000 iterations at s=2 successfully elevated both p-values to the aforementioned levels. Consequently, we utilized this final checkpoint (s=2, 70,000 iterations) throughout the conditional experiments, reporting the intermediate checkpoints solely as curriculum diagnostics.

Collectively, the tokenizer diagnostics and prior-fidelity analysis establish two critical prerequisites for the subsequent fusion experiments: the sensor-specific branches effectively receive observational signals that survive latent compression, and all branches constrain the same reusable precipitation distribution.

\begin{figure}
\centering
\includegraphics[width=0.95\linewidth,height=\textheight,keepaspectratio]{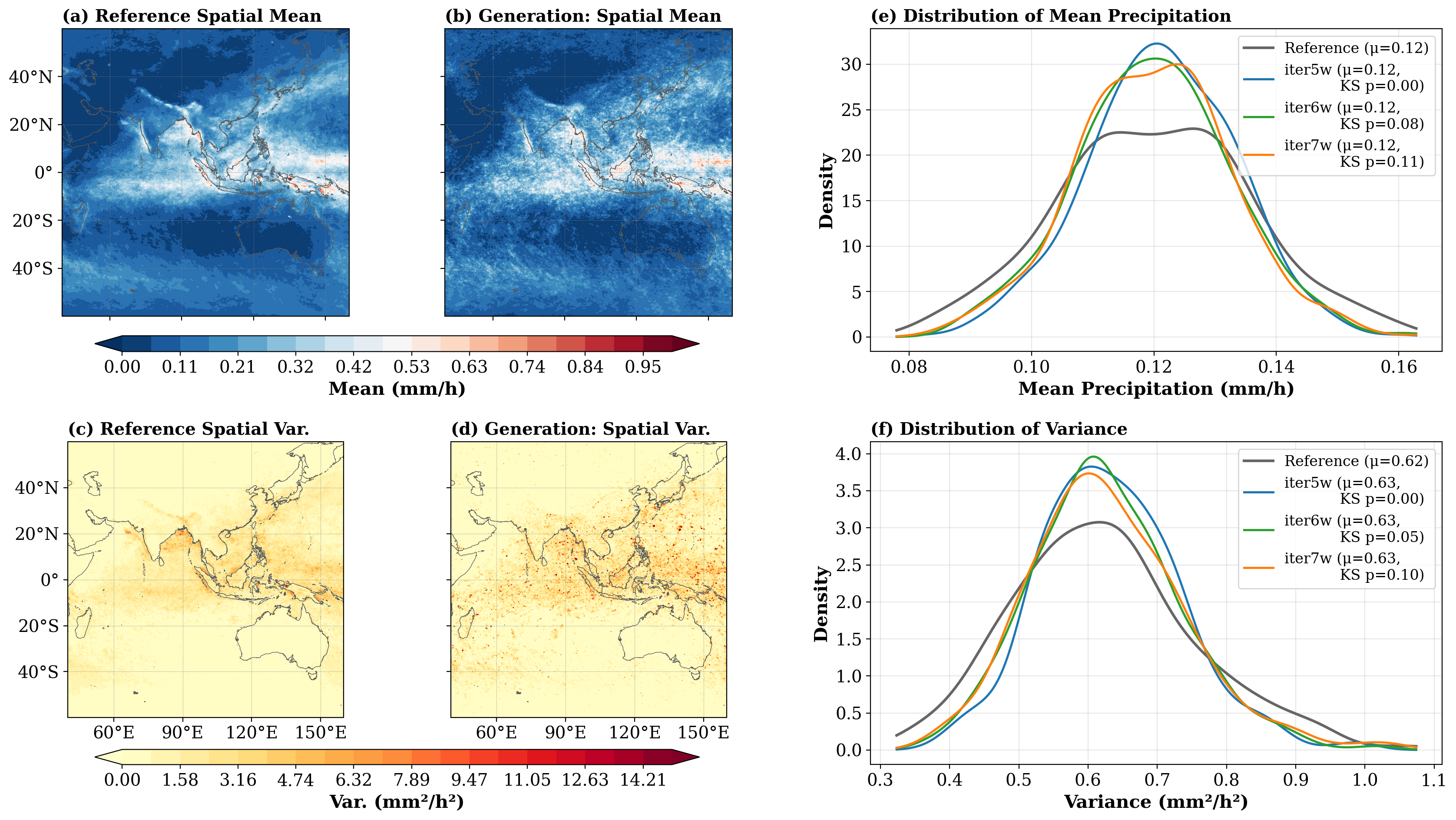}
\caption{Comparison of the spatial precipitation statistics between the reference data and generated samples. (a) and (b) show the spatial distributions of the mean precipitation rate for the reference and generated data, respectively. (c) and (d) present the corresponding spatial distributions of precipitation variance. (e) and (f) compare the probability-density distributions of the domain-mean precipitation rate and precipitation variance, respectively, for the reference data and samples generated by the iter5w, iter6w, and iter7w models. The legends report the distribution mean (\(\mu\)) and the two-sample Kolmogorov--Smirnov (KS) test \(p\)-value relative to the reference distribution.}\label{fig:prior}
\end{figure}

\subsection{Evaluation of successive sensor integration across the full domain}\label{successive-sensor-integration-yields-coverage-dependent-full-domain-gains}

We next investigated whether the independently trained sensor branches actively enhanced the frozen generative system, rather than merely maintaining architectural compatibility with it. We evaluated all three configurations at identical analysis times throughout July and August 2025, generating complete ten-member ensembles for each timestep over the 40\(\,^{\circ}\)--160\(\,^{\circ}\) E, 60\(\,^{\circ}\) S--60\(\,^{\circ}\) N domain. Here, categorical event detection was evaluated from ensemble exceedance probabilities using probability decision thresholds selected by two-fold temporal cross-validation (CV-CSI). This avoids relying solely on the ensemble mean, which can suppress event information contained in individual members, particularly for localized high-intensity precipitation. Adding GMI to the AGRI baseline increased the CV-CSI from 0.442 to 0.451, 0.366 to 0.376, 0.259 to 0.264 and 0.186 to 0.189 at the four respective thresholds. Simultaneously, this addition reduced the RMSE from 0.760 to 0.751 mm h\(^{-1}\) and the CRPS from 0.119 to 0.117 mm h\(^{-1}\) (\textbf{Fig. 4a}). Incorporating the F16--F18 SSMIS constellation yielded further enhancements, elevating the CV-CSI to 0.467, 0.395, 0.278 and 0.199 at the four respective thresholds, while further reducing the RMSE and CRPS to 0.734 and 0.113 mm h\(^{-1}\). Thus, performance improved monotonically from AGRI only to AGRI+GMI and then to AGRI+GMI+SSMIS across all reported deterministic and probabilistic metrics.

Paired comparisons showed that these aggregate improvements were systematic rather than being dominated by a small number of individual scenes. Relative to their respective baselines, the sequential addition of GMI and SSMIS improved the CV-CSI at 1 mm h\(^{-1}\) by 2.67\% and 5.01\%, respectively. The CRPS reductions were 1.99\% and 3.65\%. The 95\% confidence intervals from 5,000 bootstrap replicates blocked by UTC date remained entirely above zero for all evaluated metrics (\textbf{Fig. 4a}). Per-timestep analysis further demonstrated the broad reproducibility of these benefits (\textbf{Fig. 4b}). CV-CSI at 1 mm h\(^{-1}\) improved at 94.3\% and 88.1\% of the matched times after adding GMI and SSMIS, respectively. The corresponding proportions showing reduced CRPS were 96.0\% and 84.8\%. Although minor negative tails remained, the predominance of positive paired changes indicated that microwave constraints provided consistent incremental gains rather than improvements confined to a few high-impact events.

The magnitude of the full-domain improvement nevertheless depended strongly on the fraction of each scene directly sampled by the microwave sensors. Relative to AGRI-only, the CV-CSI gain of AGRI+GMI+SSMIS correlated positively with instantaneous GMI--SSMIS union coverage at all four thresholds (\textbf{Fig. 4c}). The correlation coefficients were 0.878, 0.826, 0.692 and 0.599 at 0.1, 1, 5 and 10 mm h\(^{-1}\), respectively. Greater coverage was also associated with larger reductions in RMSE and CRPS, with correlation coefficients of 0.722 and 0.820, respectively. Each 10-percentage-point increase in union coverage was associated with absolute CV-CSI increases of 0.016 at 0.1 mm h\(^{-1}\) and 0.020 at 1 mm h\(^{-1}\). Their corresponding 95\% confidence intervals were 0.015--0.017 and 0.018--0.022. The increases remained positive at higher thresholds, reaching 0.017 at 5 mm h\(^{-1}\) and 0.013 at 10 mm h\(^{-1}\). This coverage dependence was consistent with the modest domain-mean improvement despite the strong local information content of the microwave observations. At any analysis time, low-Earth-orbiting microwave sensors sampled only part of the geostationary domain, leaving the remaining area without direct microwave constraints. Consistent with this coverage dependence, adding DPR-Ka to AGRI+GMI+SSMIS produced nearly identical full-domain performance across the evaluated metrics (\textbf{Table S6}). This negligible domain-mean change reflects the extremely sparse instantaneous coverage of DPR-Ka rather than an absence of locally useful radar information, as examined over matched common-coverage regions in Section 4.3.

To illustrate how these coverage-dependent gains emerged spatially, we examined the analysis at 22:00 UTC on 5 July 2025 (\textbf{Fig. 4d--l}). GMI, SSMIS and DPR-Ka covered 8.3\%, 17.5\% and 2.7\% of the full domain, respectively. Despite this incomplete coverage, the four-sensor configuration increased the full-domain CV-CSI at 1 mm h\(^{-1}\) from 0.354 to 0.411. It also reduced RMSE from 0.784 to 0.730 mm h\(^{-1}\) and CRPS from 0.127 to 0.114 mm h\(^{-1}\) (\textbf{Fig. 4d--f}). The two enlarged regions showed how different microwave sensors contributed within their observed footprints. Over the northwestern Pacific, GMI sampled most of a cyclonic precipitation system, while DPR-Ka provided a narrower cross-track constraint. The multi-sensor estimate recovered its compact central structure and curved rainbands, which were broadened and displaced in the AGRI-only estimate (\textbf{Fig. 4g--i}). Within this region, CV-CSI increased from 0.421 to 0.707, while RMSE decreased from 1.008 to 0.674 mm h\(^{-1}\). CRPS concurrently decreased from 0.203 to 0.113 mm h\(^{-1}\). Over South and Southeast Asia, SSMIS covered nearly the entire enlarged region and constrained two organized precipitation bands (\textbf{Fig. 4j--l}). The multi-sensor estimate reduced their spatial overextension and better reproduced the intensity distribution in the IMERG reference. CV-CSI increased from 0.572 to 0.715, while RMSE decreased from 2.295 to 1.413 mm h\(^{-1}\). CRPS also decreased from 0.609 to 0.359 mm h\(^{-1}\). These local improvements were considerably larger than the full-domain averages, visually reinforcing that realized gains scale with the area receiving direct microwave constraints.

\begin{figure}
\centering
\includegraphics[width=0.95\linewidth,height=\textheight,keepaspectratio]{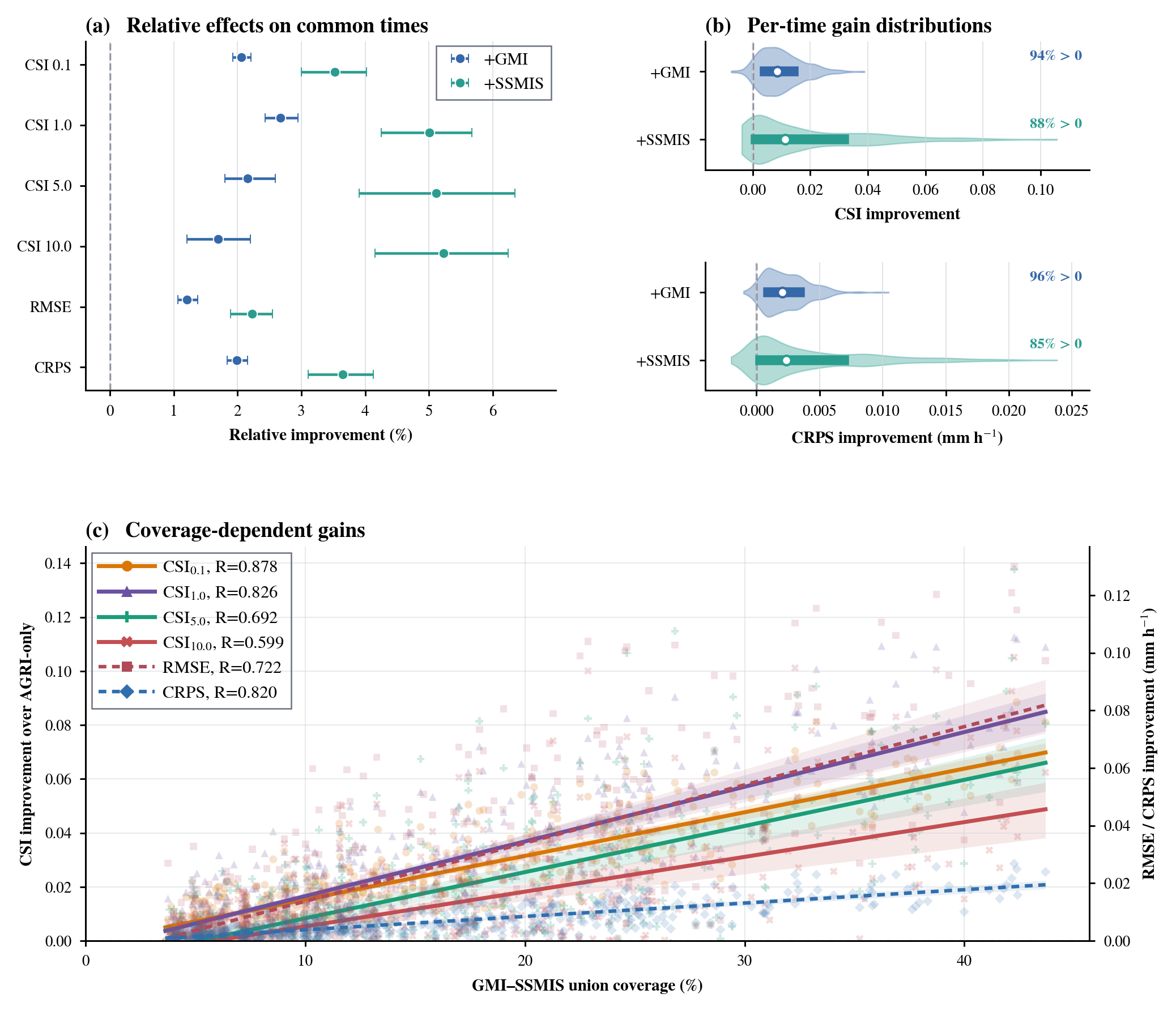}
\caption{Successive sensor integration yields consistent and coverage-dependent full-domain gains. (a), Relative improvements in CC, CV-CSI at 0.1 and 1 mm h\(^{-1}\), RMSE and CRPS after adding GMI to AGRI-only and subsequently adding SSMIS to AGRI+GMI. Points denote paired mean changes across matched analysis times, and whiskers show 95\% confidence intervals from 5,000 bootstrap resamples blocked by UTC date. (b), Distributions of per-time improvements in CV-CSI at 1 mm h\(^{-1}\) and CRPS for the two sensor additions. Thick horizontal lines span the interquartile range, white-centred points denote medians, and annotations report the percentage of times with positive improvement. (c), Per-time improvements from AGRI-only to AGRI+GMI+SSMIS as a function of GMI--SSMIS union coverage. CV-CSI improvements are shown on the left axis, while RMSE and CRPS reductions are shown on the right axis. Lines denote linear fits, shaded bands denote 95\% confidence intervals from UTC-date-block bootstrap resampling, and legend values are Pearson correlations.}\label{fig:sensor_scalability}
\end{figure}

\begin{figure}[!htbp]
\centering
\includegraphics[width=0.95\linewidth,height=\textheight,keepaspectratio]{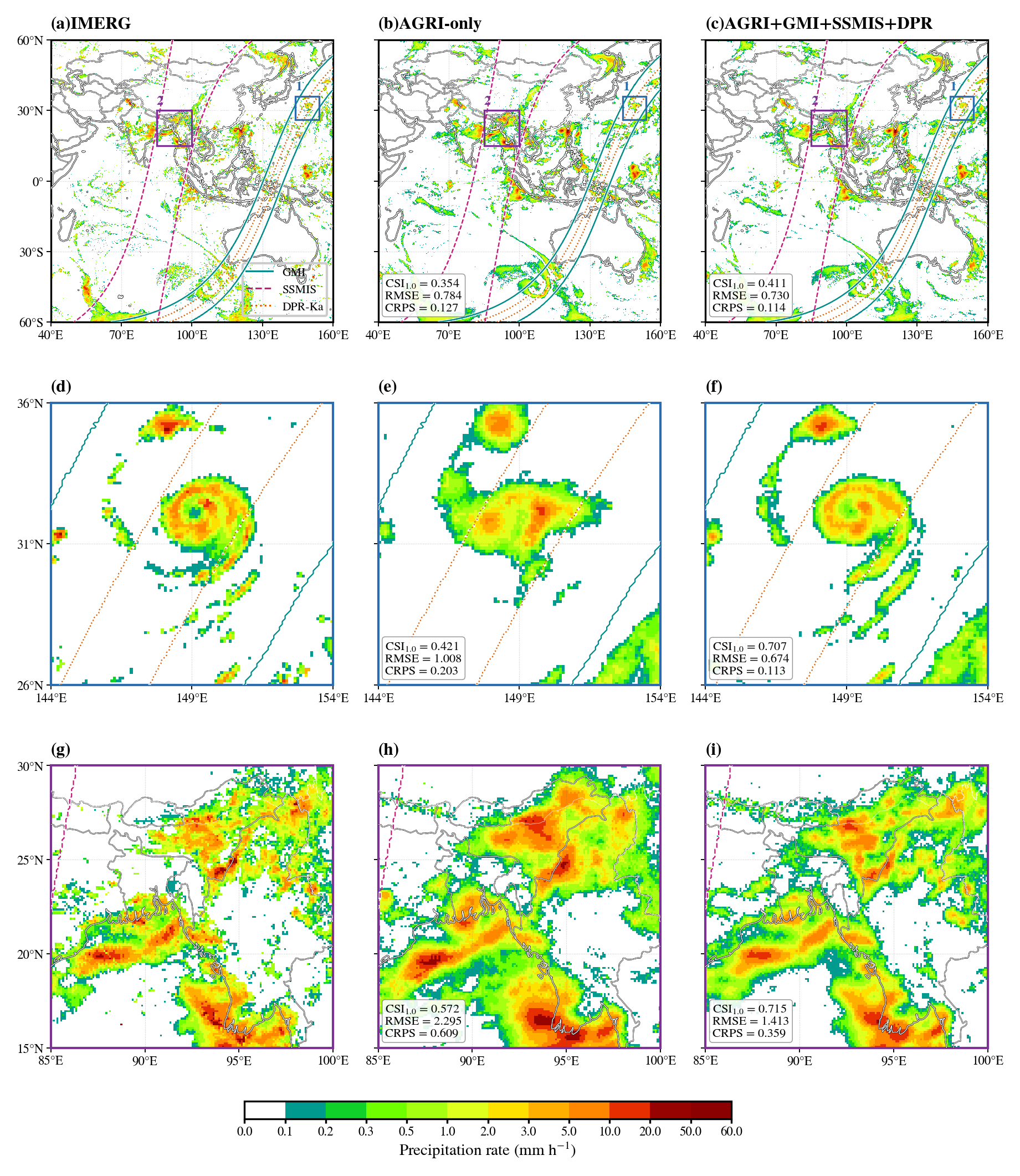}
\caption{Microwave observations yield spatially localized gains within their instantaneous coverage. Precipitation retrievals at 22:00 UTC on 5 July 2025. (a)--(c), Full-domain IMERG reference, AGRI-only estimate and AGRI+GMI+SSMIS+DPR-Ka estimate, respectively. Outlines indicate the GMI, SSMIS and DPR-Ka footprints, and numbered boxes mark the enlarged regions. (d)--(f), Corresponding views over the northwestern Pacific (144\(\,^{\circ}\)--154\(\,^{\circ}\) E, 26\(\,^{\circ}\)--36\(\,^{\circ}\) N), where GMI and DPR-Ka sampled a cyclonic precipitation system. (g)--(i), Corresponding views over South and Southeast Asia (85\(\,^{\circ}\)--100\(\,^{\circ}\) E, 15\(\,^{\circ}\)--30\(\,^{\circ}\) N), where SSMIS sampled two organized precipitation bands. Retrieval maps show ten-member ensemble means.}\label{fig:coverage_case}
\end{figure}
\FloatBarrier

\subsection{Evaluation of multi-sensor fusion over common-coverage regions}\label{multi-sensor-fusion-provides-complementary-skill-beyond-single-sensor-retrievals}

Full-domain gains from adding a sensor can reflect either an expansion of the observed area or an improvement in retrieval skill where multiple sensors observe the same region. To isolate this synergistic effect, we evaluated each fusion configuration against its stronger single-sensor baseline over strictly matched analysis times, common sensor footprints (\textbf{Fig. 6}). Specifically, the GMI--SSMIS comparison was restricted solely to their intersecting footprints, and the GMI--DPR-Ka evaluation was similarly constrained.

Across matched GMI--SSMIS analysis times, the fused estimate consistently outperformed the SSMIS-only baseline across all evaluated metrics. The CV-CSI increased from 0.613 to 0.621, 0.571 to 0.588, 0.404 to 0.417 and 0.316 to 0.324 at the four respective thresholds. RMSE decreased from 0.411 to 0.389 mm h\(^{-1}\), while CRPS decreased from 0.080 to 0.076 mm h\(^{-1}\). The 95\% UTC-date block-bootstrap intervals remained above zero for CV-CSI at 0.1, 1 and 5 mm h\(^{-1}\), RMSE and CRPS (\textbf{Fig. 6a}). The interval for CV-CSI at 10 mm h\(^{-1}\) crossed zero (-0.89\% to 5.87\%). These results support an incremental benefit from combining GMI and SSMIS beyond the stronger single-sensor baseline, although the improvement at the highest threshold was less precisely constrained.

Similarly, the GMI--DPR-Ka experiment provided an independent test incorporating active radar information. Relative to GMI-only, fusion increased CV-CSI from 0.604 to 0.609, 0.575 to 0.598, 0.438 to 0.449 and 0.328 to 0.333 at the four respective thresholds. RMSE decreased from 0.729 to 0.705 mm h\(^{-1}\), while CRPS decreased from 0.126 to 0.121 mm h\(^{-1}\). The corresponding relative CV-CSI improvements at 0.1, 1, 5 and 10 mm h\(^{-1}\) were 0.85\%, 4.00\%, 2.55\% and 1.45\%, respectively. RMSE and CRPS were reduced by 3.27\% and 4.36\%, respectively (\textbf{Fig. 6b}). The 95\% UTC-date block-bootstrap intervals remained above zero for CV-CSI at 0.1, 1 and 5 mm h\(^{-1}\), RMSE and CRPS. In contrast, the interval for CV-CSI at 10 mm h\(^{-1}\) crossed zero (-1.59\% to 4.51\%).

To illustrate how these aggregate statistical gains physically reshape the generated precipitation fields, we analyzed representative spatial examples. In the GMI--SSMIS case at 21:00 UTC on 8 July 2025 (\textbf{Fig. 6c--f}), the intersection sampled a chain of deep convective systems. Compared with either single-sensor retrieval, the fused estimate more closely reproduced the IMERG reference by suppressing spurious precipitation while recovering convective cores that were weak or absent in the individual retrievals. Combining the two sensors increased CV-CSI to 0.763 and reduced RMSE and CRPS to 1.509 and 0.506 mm h\(^{-1}\). A similar rectifying effect was observed in the GMI--DPR-Ka case at 20:30 UTC on 25 July 2025 (\textbf{Fig. 6g--j}) along the southern flank of an organized precipitation system over central India. Here, the active radar constraint successfully suppressed a broad area of spurious precipitation where the reference indicated dry conditions. This correction reduced the southward overextension of the retrieved system and sharpened its boundary with the surrounding rain-free region, elevating the CV-CSI to 0.520 and reducing RMSE from 0.486 to 0.475 mm h\(^{-1}\) and CRPS from 0.102 to 0.093 mm h\(^{-1}\) relative to GMI-only.

\begin{figure}[!htbp]
\centering
\includegraphics[width=0.95\linewidth,height=\textheight,keepaspectratio]{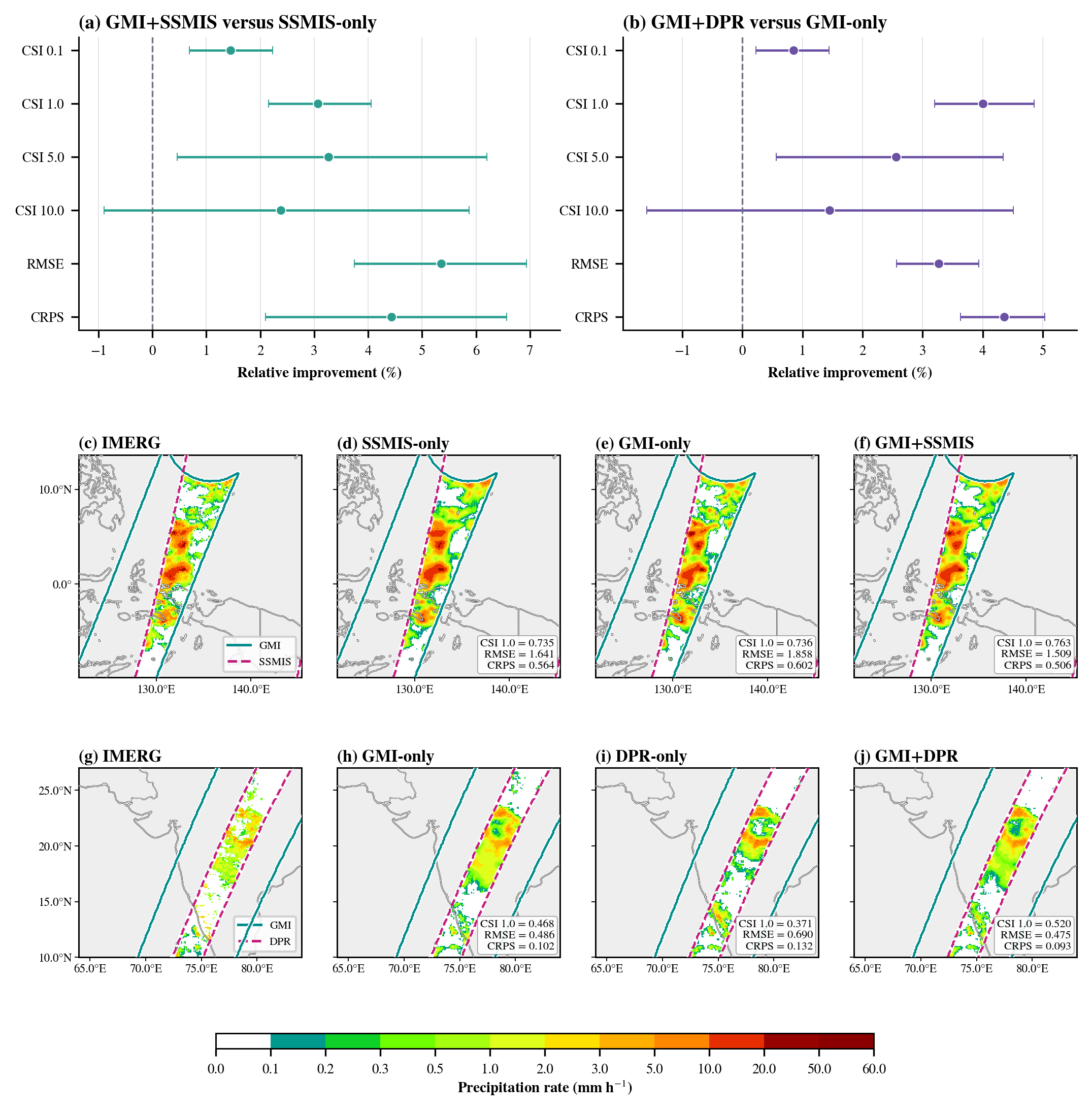}
\caption{Multi-sensor fusion provides complementary skill beyond single-sensor retrievals. (a) Relative improvements in CV-CSI at 0.1, 1.0, 5.0 and 10.0 mm h\(^{-1}\), RMSE and CRPS for GMI+SSMIS fusion relative to SSMIS-only and (b) GMI+DPR-Ka fusion relative to GMI-only. All configurations are evaluated over identical times, common sensor-coverage pixels. Points denote paired mean changes across matched analysis times, and whiskers show 95\% confidence intervals from 5,000 bootstrap resamples blocked by UTC date. (c)--(f), IMERG reference, SSMIS-only, GMI-only and GMI+SSMIS retrievals, respectively, for the common GMI--SSMIS footprint at 21:00 UTC on 8 July 2025. (g)--(j), IMERG reference, GMI-only, DPR-Ka-only and GMI+DPR-Ka retrievals, respectively, for the common GMI--DPR-Ka footprint at 20:30 UTC on 25 July 2025.}\label{fig:fusion}
\end{figure}
\FloatBarrier

\subsection{Independent gauge validation}\label{independent-gauge-validation}

Because the preceding evaluations use IMERG as the reference and IMERG also provides the training target for the precipitation prior, agreement with IMERG alone does not constitute a fully independent assessment of retrieval accuracy. We therefore evaluated PRISMA against hourly precipitation accumulations from China's national automatic weather station network. These station observations were not used as model inputs or training labels for the tokenizer, precipitation backbone, or sensor-specific conditional branches. The evaluation covers July and August 2025 and uses strict microwave-coverage samples: GMI and SSMIS are each required to provide valid observations at both half-hourly time steps contributing to a 1-h accumulation, and the resulting GMI- and SSMIS-coverage samples are pooled using station--hour sample-size weighting. This yields approximately 3.52 million station--hour pairs across 204 valid hourly samples. All station-based metrics use point-to-point collocation with the nearest \(0.1^{\circ}\) grid cell. PRISMA is evaluated using ten ensemble members, while IMERG Final, IMERG Early, PERSIANN, PERSIANN-CCS and FY-4B AGRI QPE are evaluated on the same pooled station samples and with the same spatiotemporal collocation protocol (\textbf{Fig. 7}).

The improvement was first evident in deterministic precipitation accuracy (\textbf{Fig. 7a}). Using the ensemble mean as the deterministic estimate, RMSE decreased progressively from 1.598 mm for AGRI only to 1.585 mm for AGRI+GMI, 1.545 mm for AGRI+GMI+SSMIS, and 1.541 mm for the four-source configuration. The latter was 3.7\% lower than IMERG Final (1.600 mm) and 4.7\% lower than IMERG Early (1.617 mm). Thus, the incremental improvement with successive sensor integration observed against IMERG was reproduced against an independent ground reference. The advantage was substantially more pronounced in probabilistic accuracy (\textbf{Fig. 7b}). CRPS decreased from 0.243 mm for AGRI only to 0.235 mm for AGRI+GMI and 0.227 mm for AGRI+GMI+SSMIS, and further to 0.226 mm when DPR-Ka was included. The four-source configuration therefore reduced CRPS by 29.8\% relative to IMERG Final (0.321 mm) and by 27.6\% relative to IMERG Early (0.312 mm), with reductions of 36.1--39.9\% relative to the three infrared-based products. For deterministic products, CRPS reduces to MAE because the predictive distribution is represented by a point mass, placing the ensemble and deterministic products on the same absolute-error scale. The much larger separation in CRPS than in RMSE indicates that the principal advantage of PRISMA lies not only in improving the ensemble-mean precipitation estimate, but also in representing the uncertainty of the conditional precipitation distribution.

The strongest distinction from the deterministic benchmark products appeared in threshold-based probabilistic skill (\textbf{Fig. 7c}). Using the observed event frequency as the climatological reference and applying the finite-ensemble correction of Ferro \cite{ferro2014fair}, the four-source PRISMA ensemble achieved positive fair Brier skill scores (BSSs) of 0.211, 0.171, 0.078 and 0.031 at 1-h precipitation-accumulation thresholds of 0.1, 1, 5 and 10 mm, respectively. AGRI+GMI+SSMIS likewise retained positive skill at all four thresholds, with corresponding values of 0.199, 0.161, 0.076 and 0.030. By contrast, all deterministic benchmark products yielded negative BSSs across the evaluated thresholds; IMERG Final, for example, ranged from -0.408 to -0.220. This separation should not be interpreted simply as superior deterministic retrieval accuracy. A deterministic product assigns an exceedance probability of either zero or one, whereas the PRISMA ensemble represents intermediate probabilities through the fraction of members exceeding each threshold. The positive fair BSS therefore indicates that the ensemble retains useful probabilistic information about event occurrence rather than collapsing uncertainty into a single precipitation field. The four-source configuration achieved CV-CSI values of 0.394, 0.303, 0.144 and 0.078 at precipitation-accumulation thresholds of 0.1, 1, 5 and 10 mm, respectively (\textbf{Fig. 7d}). For comparison, the fixed-threshold CSI values of IMERG Final were 0.381, 0.305, 0.161 and 0.075. PRISMA therefore showed slightly higher categorical skill at 0.1 and 10 mm, essentially comparable skill at 1 mm, and lower skill at 5 mm. Together with the consistently positive fair BSS, these results show that the ensemble retains useful event information across precipitation intensities while preserving an explicit representation of retrieval uncertainty.

\begin{figure}[!htbp]
\centering
\includegraphics[width=0.98\linewidth,height=0.92\textheight,keepaspectratio]{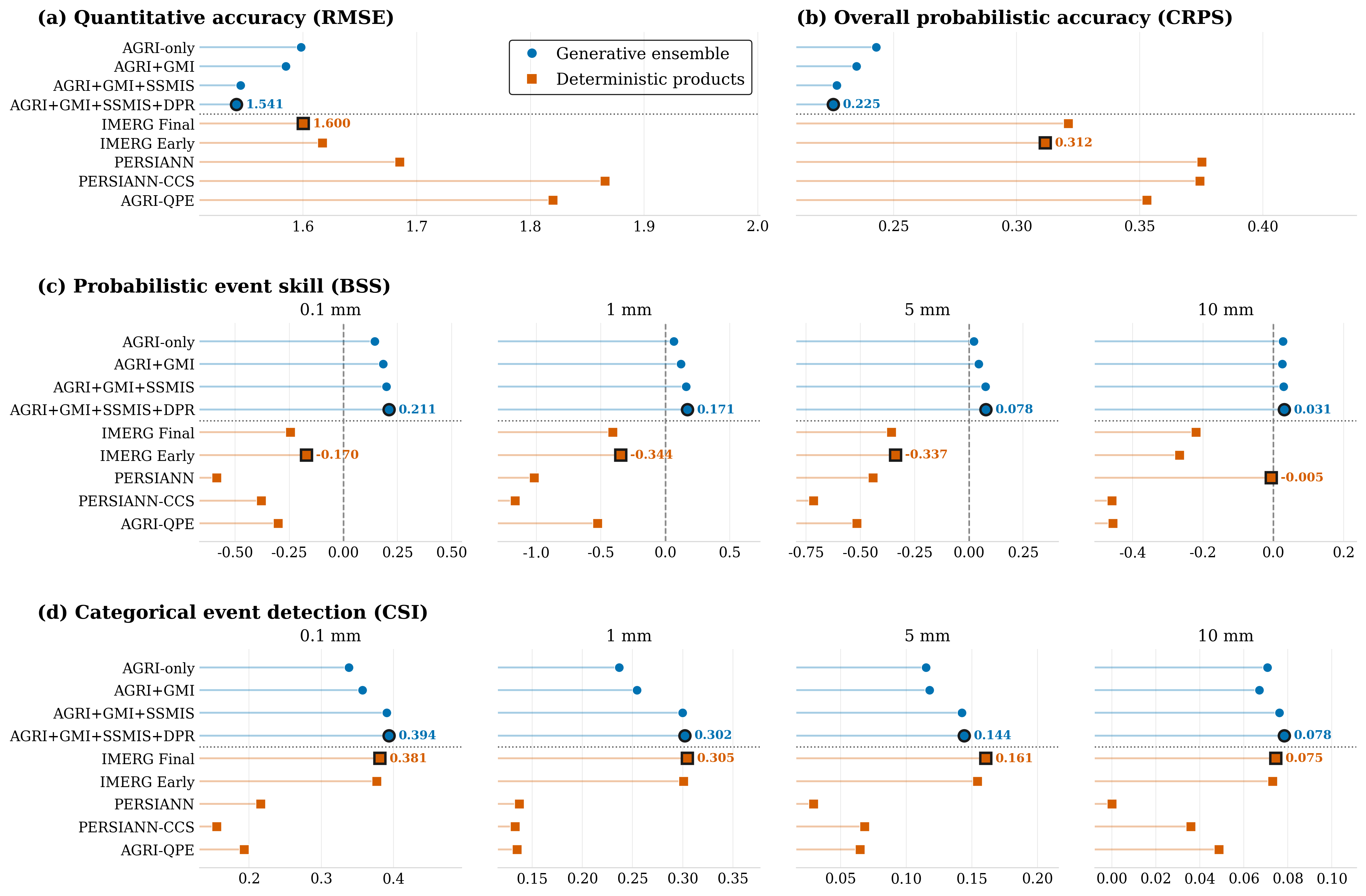}
\caption{Independent gauge evaluation across PRISMA configurations and benchmark precipitation products. (a) Quantitative precipitation accuracy measured by RMSE. (b) Overall probabilistic accuracy measured by CRPS. For PRISMA, CRPS is computed from the ten-member ensemble; for deterministic products, CRPS reduces to mean absolute error because the predictive distribution is represented by a point mass. (c) Threshold-based probabilistic skill measured by BSS at precipitation thresholds of 0.1, 1, 5 and 10 mm. A finite-ensemble correction is applied to PRISMA to obtain the fair BSS, whereas the correction vanishes for deterministic products. The vertical dashed line denotes zero skill relative to climatology. (d) Categorical event detection measured by the cross-validated critical success index (CV-CSI) at the same precipitation thresholds.}\label{fig:gauge_summary}
\end{figure}
\FloatBarrier

A representative case further illustrates these results for the 1-h accumulation ending at 21:00 UTC on 19 July 2025 (\textbf{Fig. 8}). Both the observations and retrievals show a pronounced southwest--northeast-oriented precipitation band over eastern China, together with rainfall over coastal South China and the adjacent ocean. The AGRI+GMI+SSMIS+DPR retrieval reproduces the location, orientation and spatial continuity of the primary rainband (\textbf{Fig. 8b}) relative to IMERG Final (\textbf{Fig. 8c}). Over stations within the microwave-covered region, the four-source retrieval achieved an RMSE of 1.300 mm and a CRPS of 0.250 mm, compared with 1.368 mm and 0.374 mm for IMERG Final, corresponding to reductions of approximately 5.0\% and 33.2\%, respectively. The station-level absolute-error difference further showed a mean improvement of 0.156 mm relative to IMERG Final (\textbf{Fig. 8d}). At the 1 mm threshold, CV-CSI increased from 0.409 for IMERG Final to 0.460 for the four-source retrieval, while fair BSS increased from -0.054 to 0.412. At 0.1 mm h\(^{-1}\), the CV-CSI of the four-source retrieval was slightly lower than that of IMERG Final (0.405 versus 0.426). At the 10 mm threshold, all retrievals exhibited low CV-CSI and near-zero or negative fair BSS, indicating that skill for intense precipitation remained limited in this individual case.

\begin{figure}[!htbp]
\centering
\includegraphics[width=0.95\linewidth,height=\textheight,keepaspectratio]{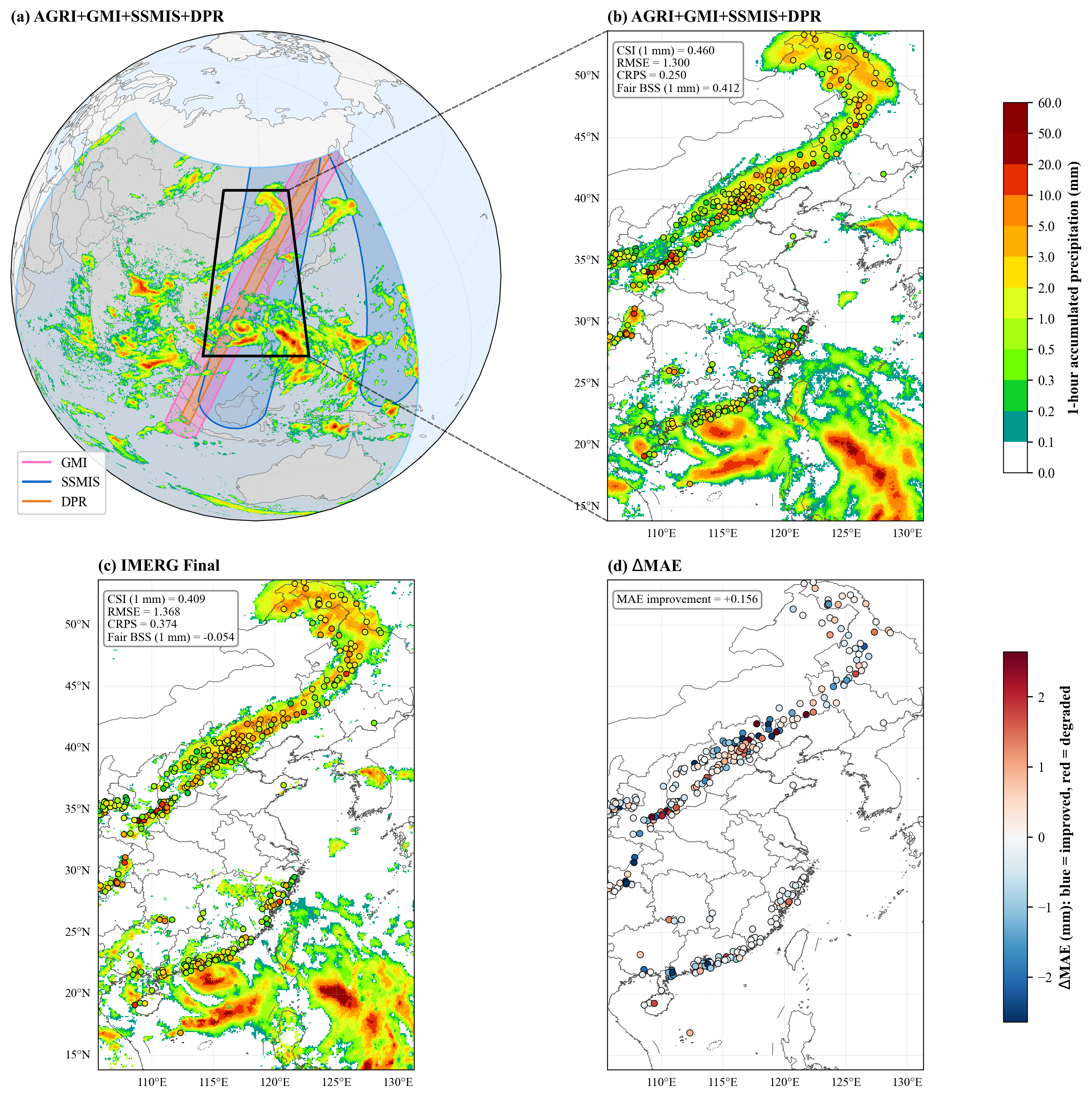}
\caption{Comparison of 1-hour accumulated precipitation ending at 21:00 UTC on 19 July 2025. (a) Global view of the AGRI+GMI+SSMIS+DPR ensemble-mean retrieval. The magenta, blue, and orange regions denote the strict two-slot footprints of GMI, SSMIS, and DPR, respectively, based on the intersection of the 20:00 and 20:30 UTC observations. The black rectangle indicates the regional domain shown in the remaining panels. (b) Regional AGRI+GMI+SSMIS+DPR ensemble-mean retrieval and (c) IMERG Final precipitation. Colored circles in (b) and (c) represent 1-hour accumulated precipitation measured by rainy national AWS stations. (d) Station-level difference in absolute error, defined as \(\Delta\)MAE = \textbar AGRI+GMI+SSMIS+DPR - AWS\textbar{} - \textbar IMERG Final - AWS\textbar. The MAE improvement, defined as the IMERG Final absolute error minus the four-source absolute error, is 0.156.}\label{fig:gauge_case}
\end{figure}
\FloatBarrier

\section{Discussion}\label{discussion}

In this study, we present PRISMA, a composable latent generative framework for precipitation estimation from multi-source satellite observations. PRISMA consists of domain-adapted tokenizers for information compression, an unconditional generative backbone that learns the precipitation distribution in latent space, and multiple conditional branches that incorporate sensor-specific observations. By separating the conditional branches from the generative backbone, PRISMA allows new observation types to be integrated through independently trained branches without retraining the precipitation backbone, providing a flexible way to accommodate continuously evolving satellite observing systems.

We demonstrate these capabilities by integrating FY-4B AGRI, GPM GMI, the F16--F18 SSMIS constellation, and GPM DPR-Ka within PRISMA. First, domain adaptation enables the tokenizers to efficiently compress both precipitation fields and heterogeneous satellite observations while retaining the information required for subsequent generative modeling, reducing the computational burden of both training and inference. Second, successive integration of microwave observations consistently improves deterministic and probabilistic precipitation estimation across the full domain, with the magnitude of the improvement increasing with instantaneous microwave coverage. Third, matched-footprint experiments show that PRISMA can effectively combine complementary information from coincident sensors and improve precipitation estimates within their overlapping footprints. Finally, independent rain-gauge validation shows that PRISMA outperforms several benchmark satellite precipitation products and achieves quantitative and categorical performance comparable to or better than IMERG Final. Together, these results show that PRISMA can support rapid and accurate precipitation estimation while explicitly representing retrieval uncertainty, providing a useful contribution to more reliable multi-satellite precipitation monitoring.

Several limitations remain in the current implementation of PRISMA. First, the spatial weighting of multiple observations is still based on observation availability and predefined sensor priorities, which may not optimally represent their relative information content. Future work could learn spatial observation weights from viewing geometry, calibration uncertainty, channel configuration, precipitation regime, and observation time, allowing the relative contribution of each sensor to vary with observing conditions. Second, the precipitation prior is trained using IMERG Final and may therefore inherit biases and smoothing characteristics of the training product. This dependence could be reduced by further fine-tuning the IMERG-trained precipitation prior with higher-fidelity references, such as radar-based precipitation products and ground-based rain-gauge observations. Third, heterogeneous satellite observations differ substantially in their native spatial resolution and sampling characteristics, while the current framework does not yet fully exploit information at their original resolutions. Developing multi-resolution observation representations and fusion strategies will be important for retaining more of the information contained in the original observations.

Looking ahead, PRISMA can be extended in two directions. One is to incorporate additional satellite observations, especially sensors that are more directly sensitive to precipitation, to further improve retrieval accuracy and robustness. The other is to generalize the framework beyond precipitation to other Earth-system state-estimation problems. A particularly promising direction is to explore flexible generative frameworks for data assimilation. Together, these directions highlight the broader potential of PRISMA as a flexible generative framework for integrating heterogeneous observations and advancing Earth-system state estimation.

\section{Methods}\label{methods}

\subsection{Problem formulation}\label{problem-formulation}

Let \(x \in \mathbb{R}^{H \times W}\) denote the target precipitation field on a regular spatial grid, and let

\begin{equation}
\mathcal{Y}=\{y^{(1)},y^{(2)},\ldots,y^{(M)}\}
\end{equation}
denote the set of available satellite observations. Each observation \(y^{(m)}\) may differ in spatial coverage, spatial resolution, sampling time, and observed variables. In particular, geostationary infrared observations provide spatially continuous but indirect information on precipitation, whereas passive-microwave and radar observations provide more direct constraints but only over intermittent orbital swaths. The available observation set therefore varies across space and time, and no individual sensor fully constrains the target precipitation field.

Our objective is to generate spatially continuous precipitation realizations that are consistent with the available observations by sampling from

\begin{equation}
x \sim p_{\theta}(x\mid\mathcal{Y}),
\end{equation}
where the conditional distribution is represented implicitly by a latent-space generative model. An overview of the PRISMA framework is shown in Fig.~\ref{fig:framework}.

PRISMA is trained in two stages (Fig.~\ref{fig:framework}b). In Stage~1, a precipitation tokenizer is adapted to the target domain and a Diffusion Transformer (DiT) parameterized with Rectified Flow is trained on IMERG precipitation fields to learn an unconditional latent precipitation prior. In Stage~2, each satellite source is equipped with an instrument-specific tokenizer and conditional branch. The precipitation backbone is frozen during this stage, and each branch is trained independently to inject sensor-specific information into the shared generative prior. At inference, any subset of the trained branches can be activated and combined through spatially varying feature weights determined by observation availability, without additional joint training. Details of tokenizer adaptation, precipitation-prior training, ControlNet-style conditional branches and their training, and multi-sensor composition and sampling are provided in Sections~\ref{latent-tokenization-and-domain-adaptation}, \ref{generative-precipitation-prior}, \ref{instrument-specific-conditional-branches}--\ref{conditional-branch-training}, and \ref{posterior-motivated-multi-sensor-composition-and-sampling}, respectively.

\subsection{Latent tokenization and domain adaptation}\label{latent-tokenization-and-domain-adaptation}

\subsubsection{Latent tokenization}\label{latent-tokenization}

Given a precipitation field \(x\), the precipitation tokenizer maps the gridded field to a latent representation

\begin{equation}
z^{\mathrm{prec}}=\mathcal{E}_{\mathrm{prec}}(x),\qquad
z^{\mathrm{prec}}\in\mathbb{R}^{C_z\times H_z\times W_z},
\end{equation}
where \(C_z\) denotes the number of latent channels and \(H_z\) and \(W_z\) denote the spatial dimensions of the latent representation. The corresponding decoder reconstructs the physical-space field as \(\hat{x}=\mathcal{D}_{\mathrm{prec}}(z^{\mathrm{prec}})\). Each satellite observation is encoded by an instrument-specific tokenizer,

\begin{equation}
z^{(m)}=\mathcal{E}_{\mathrm{sat}}^{(m)}\!\left(y^{(m)}\right).
\end{equation}
where \(m\) indexes the satellite instrument. The generative backbone and all sensor-conditioning operations are performed in this latent space, while evaluation is conducted after decoding the generated latent representation back to physical precipitation units.

\subsubsection{Tokenizer domain adaptation}\label{tokenizer-domain-adaptation}

We adapt pretrained tokenizers from the Cosmos-Tokenize1 family \cite{agarwal2024cosmos} to precipitation and satellite-observation data. Cosmos-Tokenize1-CI8x8, pretrained on natural images, is used for single-time fields, including precipitation and low-Earth-orbiting microwave and radar observations. Cosmos-Tokenize1-CV4x8x8, pretrained on natural videos, is used for sequences of geostationary AGRI observations spanning multiple consecutive times. Both tokenizers produce continuous latent representations with 16 channels and provide an \(8\times8\) spatial compression factor, while the video tokenizer additionally compresses the temporal dimension by a factor of four. Accordingly, for a \(1{,}200\times1{,}200\) single-time field, the latent representation defined above has \(C_z=16\) and \(H_z=W_z=150\).

Because the pretrained tokenizers were designed for three-channel natural imagery, their input layers are expanded when the geophysical input contains a different number of channels. Following copy-based initialization strategies previously used to transfer RGB-pretrained networks to multispectral Earth-observation data \cite{pan2019coinnet,stewart2023ssl4eo}, newly introduced channel weights are initialized by cyclically copying weights from the pretrained input channels, after which all tokenizer parameters are fine-tuned on the corresponding geophysical data. For satellite observations with incomplete spatial coverage, a binary validity mask is supplied as an additional input channel to explicitly distinguish missing observations from physically low-valued measurements, following mask-aware representations commonly used for incomplete spatial data and weather-field reconstruction \cite{schmutz2024enhanced}.

The tokenizers were fine-tuned using AdamW with a peak learning rate of \(5\times10^{-5}\)~\cite{loshchilov2019adamw}, reached after 10{,}000 warmup steps and held constant thereafter. The training objective combined a pixel-wise \(L_1\) reconstruction loss with weight 1.5 and a weak KL regularization term with weight \(10^{-6}\). For the AGRI tokenizer, an additional temporal-consistency term is activated after the first 5{,}000 warmup steps to preserve coherent evolution across consecutive AGRI observations.

\subsection{Generative precipitation prior}\label{generative-precipitation-prior}

\subsubsection{Rectified Flow with a Diffusion Transformer}\label{rectified-flow-with-a-diffusion-transformer}

The unconditional precipitation distribution is learned in latent space using Rectified Flow \cite{liu2023flow}. Let \(z_0\) denote the latent representation of a clean precipitation field obtained from the precipitation tokenizer and let \(\epsilon\sim\mathcal{N}(0,I)\) denote a latent variable independently sampled from a standard Gaussian distribution. For time variable \(t\in[0,1]\), we define the linear interpolation path

\begin{equation}
z_t=(1-t)z_0+t\epsilon,
\label{eq:rf_path}
\end{equation}
where \(z_t\) is the intermediate latent state at time \(t\). Accordingly, \(z_t=z_0\) at \(t=0\), corresponding to the precipitation latent distribution, and \(z_t=\epsilon\) at \(t=1\), corresponding to the Gaussian reference distribution. 

The evolution of the latent state along this path is represented by a learnable velocity field \(v_{\theta}(z_t,t)\), where \(v_{\theta}\) predicts the instantaneous transport direction at state \(z_t\) and time \(t\), and \(\theta\) denotes the trainable model parameters. The corresponding continuous-time dynamics are written as

\begin{equation}
\frac{d z_t}{dt}=v_{\theta}(z_t,t).
\end{equation}
We parameterize \(v_{\theta}\) using a Diffusion Transformer (DiT)~\cite{peebles2023scalable}, which models interactions among the latent tokens through self-attention. For the linear interpolation path in Eq.~\eqref{eq:rf_path}, differentiation with respect to \(t\) gives the target velocity.

\begin{equation}
\frac{d z_t}{dt}=\epsilon-z_0.
\end{equation}

The DiT is therefore trained to predict this target velocity from the intermediate latent state \(z_t\) and its associated time \(t\). Following the flow-matching objective \cite{liu2023flow,lipman2023flow}, we minimize

\begin{equation}
\mathcal{L}_{\mathrm{FM}}
=
\mathbb{E}_{z_0,\epsilon,t}
\left[
\left\|v_{\theta}(z_t,t)-(\epsilon-z_0)\right\|_2^2
\right].
\label{eq:flow_matching}
\end{equation}
Minimizing \(\mathcal{L}_{\mathrm{FM}}\) trains the DiT to predict the velocity field that transports latent states between the precipitation distribution and the Gaussian reference distribution.

\subsubsection{Time sampling and curriculum training}\label{time-sampling-and-curriculum-training}

Training timesteps are sampled from a logit-normal distribution, following common practice in diffusion-transformer training \cite{esser2024scaling}. Specifically, we sample \(u\sim\mathcal{N}(0,1)\) and map it to \((0,1)\) through the logistic sigmoid, \(t=\operatorname{sigmoid}(u)\). A shift factor \(s\) is then applied to obtain the effective training timesteps:

\begin{equation}
\tilde{t}=\frac{s t}{1+(s-1)t},
\label{eq:time_shift}
\end{equation}
where \(s>1\) increases the sampling density toward higher-noise states. We employ a two-stage curriculum \cite{sun2026curriculum,kim2025denoisingtaskdifficultybasedcurriculum}. The first 50{,}000 optimization steps use \(s=5\) to emphasize large-scale organization under relatively high noise, followed by 20{,}000 steps with \(s=2\) to place greater emphasis on lower-noise states and refine fine-scale structure and the heavy-tailed precipitation-intensity distribution. Throughout both stages, the backbone is optimized using AdamW with \(\beta_1=0.9\), \(\beta_2=0.999\)~\cite{loshchilov2019adamw}, and a weight decay of \(10^{-3}\). The effective learning rate is linearly warmed up over the first 1{,}000 optimization steps to a peak value of \(4.3\times10^{-5}\), followed by linear decay. The optimizer and scheduler states are retained when transitioning to the second stage; therefore, no additional warm-up is applied.

\subsection{Multi-source conditional generation}\label{multi-source-conditional-generation}

\subsubsection{Instrument-specific conditional branches}\label{instrument-specific-conditional-branches}

PRISMA introduces satellite constraints through independently trainable conditional branches inspired by ControlNet \cite{zhang2023controlnet}. Each instrument is assigned a branch with the same sequence of transformer blocks as the precipitation backbone. The branch processes the encoded observation for that instrument and produces layer-wise conditioning features that are mapped through zero-initialized output projections and injected additively into the corresponding blocks of the frozen backbone. In this way, each branch learns how observations from a specific satellite instrument constrains the common precipitation prior while remaining independent of the other sensor branches. This composable construction allows new sensors to be introduced without modifying the precipitation backbone or retraining previously learned branches.

\subsubsection{Conditional-branch training}\label{conditional-branch-training}

Each conditional branch is trained independently while all precipitation-backbone parameters remain frozen. Only the instrument-specific observation embedder, control blocks, and their output projections are updated. The branch is optimized with the same Rectified Flow target used for the unconditional backbone. A binary validity mask is supplied together with each satellite observation so that the branch can distinguish locations without measurements from locations containing valid low-valued observations.

The control embedder and transformer blocks are initialized from the corresponding pretrained backbone layers. The layer-wise output projections that inject branch features into the backbone are initialized exactly to zero, following the ControlNet zero-initialization principle \cite{zhang2023controlnet}. Consequently, each newly initialized branch contributes no conditioning signal at the beginning of training, and the model initially recovers the frozen unconditional precipitation prior.

Trainable parameters are optimized with AdamW \cite{loshchilov2019adamw} using \(\beta_1=0.9\), \(\beta_2=0.999\), weight decay \(10^{-3}\), and a peak learning rate of \(4.3\times10^{-5}\). The learning rate is linearly warmed up for 100 steps and then decayed with a cosine schedule \cite{loshchilov2017sgdr}. Gradient norm is clipped at 0.1. Each sensor branch is trained for 20{,}000 optimization steps using paired 2024 satellite observations and IMERG targets. Training uses fully sharded data parallelism (FSDP), context parallelism \cite{liu2023ring}, and activation checkpointing \cite{chen2016training} to reduce memory consumption for long latent sequences. Because the backbone is frozen, adding a new sensor requires optimization only of its tokenizer adaptation and conditional branch rather than retraining the precipitation prior or the previously trained sensor branches.

\subsubsection{Posterior-motivated multi-sensor composition and sampling}\label{posterior-motivated-multi-sensor-composition-and-sampling}

The composition of independently trained sensor branches is motivated by Bayesian posterior inference. Given a precipitation prior \(p(x)\) and a set of available observations \(\mathcal{Y}=\{y^{(m)}\}_m\) with sensor-specific spatial supports \(\Omega^{(m)}\), a conditionally independent observation model would yield the formal posterior:

\begin{equation}
p(x\mid\mathcal{Y})
\propto
p(x)\prod_m p\!\left(y^{(m)}\mid x_{\Omega^{(m)}}\right).
\label{eq:posterior}
\end{equation}
Because the set of available observations varies spatially and temporally, branch composition is modulated using spatial weight maps derived from the corresponding observation masks. For sensor \(m\) and spatial location \(s\), the validity mask \(M^{(m)}(s)\in\{0,1\}\) records whether a measurement is available. The feature-injection weight \(W^{(m)}(s)\) is zero outside this valid support, preventing a sensor branch from imposing a local constraint where no observation exists. The relative weighting of infrared and microwave constraints was determined empirically through a sensitivity experiment over GMI coverage regions (Table~S7). Increasing the GMI branch weight improved overall estimation accuracy, with the best performance obtained at \(g_w=1\). Accordingly, within regions covered by microwave observations, the AGRI branch weight is set to zero. When multiple microwave instruments observe the same location, equal weights are assigned to the microwave branches and normalized to sum to one. Where AGRI is the only available observation, its weight is set to one. The resulting conditional feature at transformer layer \(i\) is

\begin{equation}
h_i^{\mathrm{cond}}(s)
=
\sum_{m\in\mathcal{A}} W^{(m)}(s)\odot h_i^{(m)}(s),
\qquad
\sum_{m\in\mathcal{A}} W^{(m)}(s)\le 1,
\label{eq:fusion}
\end{equation}
where \(\mathcal{A}\) denotes the set of active sensor branches and \(h_i^{(m)}(s)\) is the feature produced by branch \(m\). The frozen backbone state is then updated as

\begin{equation}
h_i(s)
=
h_i^{\mathrm{backbone}}(s)+h_i^{\mathrm{cond}}(s).
\label{eq:branch_injection}
\end{equation}
Sampling starts from a Gaussian latent \(z_1\sim\mathcal{N}(0,I)\) and integrates the conditional Rectified Flow trajectory from \(t=1\) to \(t=0\). Using an Euler solver,

\begin{equation}
z_{t-\Delta t}
=
z_t-\Delta t\,\hat{v}_{\theta}\!\left(z_t,t,\mathcal{Y}\right),
\label{eq:euler}
\end{equation}
where \(\hat{v}_{\theta}\) is the velocity predicted by the frozen backbone after injection of the spatially composed conditional features. Because branch activation and composition require no additional parameter optimization, arbitrary subsets of trained sensors can be used at inference according to the observations available at a given analysis time. The final latent state \(z_0\) is decoded by the precipitation tokenizer to obtain a precipitation realization. Ten-member ensembles are generated using different random seeds, with each member initialized from an independently sampled \(z_1\) and generated using 35 Euler steps.

\subsection{Data}\label{data}

\subsubsection{IMERG precipitation target}\label{imerg-precipitation-target}

The study is conducted primarily over the domain spanning 60\(^\circ\)S--60\(^\circ\)N and 40\(^\circ\)--160\(^\circ\)E, with all target and observation fields mapped to a common \(0.1^\circ\) regular latitude--longitude grid of \(1{,}200\times1{,}200\) cells. This domain covers the Asian monsoon region, the Maritime Continent, the tropical western Pacific, the Indian Ocean, and the arid regions of Central and western Asia, encompassing a wide range of precipitation regimes, including tropical deep convection, monsoon rainbands, mid-latitude systems, and predominantly dry environments.

The Integrated Multi-satellitE Retrievals for GPM (IMERG) product \cite{huffman2020imerg} provides half-hourly precipitation estimates at \(0.1^\circ\) resolution by combining observations from the GPM microwave constellation with infrared information. IMERG is released as Early, Late, and Final products with progressively greater latency and additional calibration. Among them, IMERG Final includes monthly gauge adjustment and is widely used as a reference precipitation product in precipitation-related studies because of its relatively high accuracy and global coverage. Its release latency is approximately 3.5 months, which limits its use for real-time applications but makes it suitable as a high-quality gridded target for model development.

IMERG Final fields from 2021--2024 are used to adapt the precipitation tokenizer and train the unconditional generative backbone. Sensor-specific tokenizer adaptation and conditional-branch training use paired satellite observations and IMERG targets from 2024. July and August 2025 are held out for evaluation.

Precipitation rates are non-negative, strongly intermittent, and heavy-tailed. Before tokenization, precipitation values are therefore log-transformed and linearly rescaled to the \([-1,1]\) range. The transformation is inverted before all physical-space evaluation, and all reported precipitation metrics are computed in the original units.

\subsubsection{Satellite observations}\label{satellite-observations}

We use observations from one geostationary infrared imager, two passive-microwave observing systems, and one spaceborne precipitation radar. All satellite observations are mapped onto the common 0.1° IMERG grid using nearest-neighbour remapping.

The Advanced Geostationary Radiation Imager (AGRI) aboard FY-4B, located at 105\(^\circ\)E, provides the geostationary observations \cite{yang2017fy4a}. AGRI has 15 spectral channels, but visible and near-infrared measurements are not continuously available at night. We therefore use the nine infrared channels CH07--CH15 (3.75--13.30 \(\mu\)m) to maintain a consistent all-day input configuration. These include water-vapour absorption channels near 6.25, 6.95, and 7.42 \(\mu\)m, thermal-window channels that provide cloud-top brightness-temperature information, and the 13.30-\(\mu\)m CO\(_2\) absorption channel. AGRI observations are available every 15 min over the FY-4B geostationary disk.

Passive-microwave information is first provided by the GPM Microwave Imager (GMI) aboard the GPM Core Observatory \cite{hou2014gpm}. GMI is a conically scanning radiometer with 13 channels spanning 10.65--183.31 GHz and an approximately 885-km swath \cite{yang2024gmi3d}. Lower-frequency channels are primarily sensitive to microwave emission associated with liquid hydrometeors, whereas higher-frequency channels contain strong scattering signatures from ice-phase particles in precipitating clouds \cite{kummerow1998trmm,yang2024gmi3d}. The combination provides complementary information on precipitation structure, but only along the instantaneous orbital swath.

We additionally use Special Sensor Microwave Imager/Sounder (SSMIS) observations from the F16, F17, and F18 platforms. GMI and SSMIS brightness temperatures are obtained from the GPM Level-1C common-calibrated Version 07 products, in which constellation radiometers are intercalibrated against GMI~\cite{berg2016intercalibration}. Eleven channels shared by the three SSMIS instruments are retained: 19.35 GHz (V, H), 22.235 GHz (V), 37.0 GHz (V, H), 91.655 GHz (V, H), 150.0 GHz (H), and the three 183.31-GHz water-vapour channels. Combining the three SSMIS platforms increases the temporal availability of passive-microwave coverage. When observations from multiple SSMIS platforms overlap on the common grid, they are not averaged. Consistent with temporal down-selection strategies used in multi-satellite microwave compositing \cite{wimmers2011seamless,tan2019imerg}, each grid cell is assigned to the platform whose observation time is closest to the centre of the target analysis interval. The complete channel vector at that grid cell is taken from the same platform, thereby preserving a physically consistent set of channels from one instrument, acquisition time, and viewing geometry.

Active-radar information is provided by the Ka-band Precipitation Radar (KaPR) component of the GPM Dual-frequency Precipitation Radar (DPR) \cite{seto2021dpr}. We use attenuation-corrected reflectivity (\texttt{zFactorFinal}, dBZ) from the Version 07 2AKa product \cite{meneghini2021attenuation}. Because the native radar measurement is vertically resolved, each profile is reduced to a two-dimensional composite by taking the maximum attenuation-corrected reflectivity within the column. This representation retains the strongest radar echo while allowing the DPR-Ka observations to be encoded using the same two-dimensional latent-conditioning framework as the other low-Earth-orbiting sensors.

\subsubsection{External precipitation products}\label{external-precipitation-products}

For independent station-based comparison, we include FY-4B AGRI QPE, PERSIANN, and PERSIANN-CCS as external infrared-based precipitation products. These products provide modality-relevant baselines for the AGRI-only PRISMA configuration because they derive precipitation primarily from geostationary infrared observations rather than from passive-microwave measurements.

The FY-4B AGRI QPE is an operational Level-2 precipitation product with a native 15-min temporal resolution and approximately 4-km spatial resolution near the satellite subpoint \cite{song2025huayu}. PERSIANN estimates precipitation from geostationary infrared imagery using artificial neural networks and is distributed at \(0.25^\circ\) resolution \cite{nguyen2018persiann}. PERSIANN-CCS augments the PERSIANN framework with cloud classification and provides higher-resolution precipitation estimates at \(0.04^\circ\) resolution \cite{hong2004persiannccs,nguyen2018persiann}. All external products are evaluated against AWS observations using the same temporal aggregation, spatial collocation, and station sample selection as PRISMA and IMERG.

\subsection{Evaluation protocol and metrics}\label{evaluation-protocol-and-metrics}

\subsubsection{Ground-station validation}\label{ground-station-validation}

Ground-station validation uses hourly precipitation accumulations from China's national automatic weather station (AWS) network maintained by the China Meteorological Administration. These gauge observations are not used in tokenizer adaptation, precipitation-backbone training, or sensor-branch training. Station records are screened by the CMA quality-control procedure, including climatological range, spatial-consistency, and temporal-continuity checks \cite{ren2015aws}. Only station--time pairs with valid gauge observations and valid collocated gridded estimates are retained.

PRISMA and IMERG are represented as half-hourly precipitation rates on the common \(0.1^\circ\) grid. To match the hourly AWS accumulation, the two half-hourly rates contributing to an hour ending at time \(T\) are integrated as

\begin{equation}
\hat{a}_T
=
\frac{1}{2}\hat{r}_{T-1}
+
\frac{1}{2}\hat{r}_{T-0.5}
=
\frac{\hat{r}_{T-1}+\hat{r}_{T-0.5}}{2},
\label{eq:station_temporal}
\end{equation}
where \(\hat{r}\) is expressed in mm h\(^{-1}\) and \(\hat{a}_T\) in mm. The same temporal aggregation is applied to comparison products whenever their native temporal resolution is finer than 1 h.

For station-based evaluation, each station is paired with the nearest \(0.1^\circ\times0.1^\circ\) grid cell using point-to-point collocation. The station analysis is restricted to periods for which passive-microwave observations are available. For each sensor \(m\in\{\mathrm{GMI},\mathrm{SSMIS}\}\), an hourly-valid coverage mask is formed by requiring valid sensor coverage at both half-hourly times contributing to the hourly accumulation. Only stations within each sensor's strict two-slot intersection are retained, and hours containing fewer than 1{,}000 valid stations are excluded. Metrics for the combined evaluation are recomputed after pooling the GMI and SSMIS samples, which is equivalent to station–hour sample-size weighting rather than an unweighted average of the two regional scores. 

\subsubsection{Metric definitions}\label{metric-definitions}

Let \(\hat{y}_i\) and \(y_i\) denote paired prediction and reference values for \(N\) valid samples. Depending on the experiment, a sample can represent a grid cell, a station--time pair, or a tokenizer reconstruction pixel/channel. Continuous errors are quantified using

\begin{align}
\mathrm{MAE}
&=\frac{1}{N}\sum_{i=1}^{N}|\hat{y}_i-y_i|,\\
\mathrm{RMSE}
&=\sqrt{\frac{1}{N}\sum_{i=1}^{N}(\hat{y}_i-y_i)^2},\\
\mathrm{MBE}
&=\frac{1}{N}\sum_{i=1}^{N}(\hat{y}_i-y_i).
\end{align}
The Pearson correlation coefficient is

\begin{equation}
\mathrm{CC}
=
\frac{\sum_{i=1}^{N}(\hat{y}_i-\bar{\hat{y}})(y_i-\bar{y})}
{\sqrt{\sum_{i=1}^{N}(\hat{y}_i-\bar{\hat{y}})^2}
 \sqrt{\sum_{i=1}^{N}(y_i-\bar{y})^2}}.
\end{equation}
Tokenizer reconstruction is additionally evaluated using the structural similarity index (SSIM),

\begin{equation}
\mathrm{SSIM}
=
\frac{(2\mu_{\hat{y}}\mu_y+C_1)(2\sigma_{\hat{y}y}+C_2)}
{(\mu_{\hat{y}}^2+\mu_y^2+C_1)(\sigma_{\hat{y}}^2+\sigma_y^2+C_2)},
\end{equation}
where \(\mu\), \(\sigma^2\), and \(\sigma_{\hat{y}y}\) denote local means, variances, and covariance, and \(C_1\) and \(C_2\) are stabilizing constants.

For a precipitation threshold \(\tau\), hits (\(H\)), misses (\(M\)), and false alarms (\(FA\)) are accumulated from the predicted and observed binary event fields. The critical success index is

\begin{equation}
\mathrm{CSI}=\frac{H}{H+M+FA}.
\end{equation}
For categorical evaluation of PRISMA, ensemble exceedance probabilities are used rather than thresholding the ensemble mean alone. For a physical precipitation threshold \(\tau\), the exceedance probability for verification sample \(i\) is estimated from the fraction of the \(K=10\) ensemble members exceeding that threshold:

\begin{equation}
p_i(\tau)=\frac{1}{K}\sum_{k=1}^{K}
\mathbb{I}\!\left[\hat{y}_i^{(k)}\geq\tau\right],
\qquad K=10.
\label{eq:cv_csi_probability}
\end{equation}
Because \(K=10\), the candidate probability decision thresholds are \(q\in\{0.1,0.2,\ldots,1.0\}\). Following established probability-threshold-based categorical verification and cross-validated evaluation practices \cite{roebber2009visualizing,gagne2017storm,gensini2021machine}, the decision threshold is selected using two-fold temporal cross-validation, with even and odd UTC calendar dates defining folds A and B, respectively.

For each precipitation threshold \(\tau\),

\begin{equation}
q^*(\tau)
=
\underset{q}{\operatorname{arg\,max}}\;
\mathrm{CSI}(\tau,q).
\label{eq:cv_csi_threshold}
\end{equation}
The threshold selected from fold A, \(q_A^*(\tau)\), is applied unchanged to the held-out fold B, whereas \(q_B^*(\tau)\) is applied to fold A. All valid verification samples within each held-out fold are pooled to form contingency counts, and the resulting out-of-sample counts are then combined to obtain the cross-validated CSI (hereafter, CV-CSI),

\begin{equation}
\resizebox{0.96\linewidth}{!}{$\displaystyle
\mathrm{CV\text{-}CSI}(\tau)
=
\frac{
H_B(\tau,q_A^*)+H_A(\tau,q_B^*)
}{
H_B(\tau,q_A^*)+H_A(\tau,q_B^*)
+M_B(\tau,q_A^*)+M_A(\tau,q_B^*)
+FA_B(\tau,q_A^*)+FA_A(\tau,q_B^*)
}.$}
\label{eq:cv_csi}
\end{equation}
Probabilistic accuracy of the ensemble is evaluated using the continuous ranked probability score (CRPS) \cite{gneiting2007strict,price2025gencast}. Let \(\{\hat{y}_i^{(k)}\}_{k=1}^{K}\) denote the \(K\) ensemble members for sample \(i\). The ensemble CRPS estimator is

\begin{align}
\mathrm{CRPS}
&=
\frac{1}{N}\sum_{i=1}^{N}
\left[
\frac{1}{K}\sum_{k=1}^{K}|\hat{y}_i^{(k)}-y_i|
\right.\nonumber\\
&\qquad\left.
-
\frac{1}{2K^2}\sum_{k=1}^{K}\sum_{\ell=1}^{K}
|\hat{y}_i^{(k)}-\hat{y}_i^{(\ell)}|
\right].
\end{align}
The first term measures the average distance between ensemble members and the reference, while the second term rewards ensemble spread when it appropriately represents predictive uncertainty. Lower CRPS indicates better probabilistic performance. For a deterministic product represented by a point-mass predictive distribution, CRPS reduces to absolute error, and its sample mean is therefore identical to MAE. This permits deterministic products and PRISMA ensembles to be compared on the same absolute-error scale in the station evaluation.

Threshold-specific probabilistic skill is evaluated using the finite-ensemble fair Brier score. For event threshold \(\tau\), the ensemble exceedance probability is estimated as

\begin{equation}
p_i(\tau)=\frac{k_i(\tau)}{K},
\qquad
k_i(\tau)=\sum_{k=1}^{K}
\mathbb{I}\!\left[\hat{y}_i^{(k)}\ge\tau\right],
\label{eq:ens_prob}
\end{equation}
with observed event indicator \(o_i(\tau)=\mathbb{I}[y_i\ge\tau]\). Because \(p_i\) is estimated from a finite ensemble, the conventional Brier score contains a sampling contribution that depends on ensemble size. We therefore use the finite-ensemble correction of Ferro \cite{ferro2014fair},

\begin{equation}
\mathrm{BS}_{\mathrm{fair},i}
=
(p_i-o_i)^2
-
\frac{k_i(K-k_i)}{K^2(K-1)}.
\label{eq:fair_bs}
\end{equation}
The fair Brier skill score uses the observed event frequency \(s=N^{-1}\sum_i o_i\) as the climatological reference,

\begin{equation}
\mathrm{BSS}_{\mathrm{fair}}
=
1-
\frac{N^{-1}\sum_{i=1}^{N}\mathrm{BS}_{\mathrm{fair},i}}
{s(1-s)}.
\label{eq:fair_bss}
\end{equation}
Positive values indicate improvement over the climatological event probability, zero denotes no improvement over climatology, and negative values indicate lower probabilistic skill than the climatological reference. Deterministic products are represented by probabilities of zero or one, for which the finite-ensemble correction vanishes, and are therefore evaluated using the conventional Brier score within the same skill-score formulation.

For distributional comparisons of generated and reference samples, we use the two-sample Kolmogorov--Smirnov statistic. Given empirical cumulative distribution functions \(F_n\) and \(G_m\),

\begin{equation}
D_{n,m}=\sup_x|F_n(x)-G_m(x)|,
\end{equation}
with the corresponding \(p\)-value used to assess whether the two empirical sample distributions are statistically distinguishable.

\backmatter

\section*{Declarations}

\begin{itemize}
\item Acknowledgements: This study was supported by the National Key Research and Development Program of China [grant number 2025YFE0217100] and the National Natural Science Foundation of China [grant number U2442219]. The authors also thank NVIDIA for developing and open-sourcing the Cosmos World Foundation Model Platform, whose pretrained tokenizers served as the foundation for the observation and precipitation encoders in this work.
\item Data availability: The IMERG Final Run precipitation data are available from the NASA Goddard Earth Sciences Data and Information Services Center (GES DISC) at \url{https://disc.gsfc.nasa.gov/datasets/GPM_3IMERGHH_07/summary}. GPM GMI and F16--F18 SSMIS Level-1C common-calibrated brightness temperatures and GPM DPR-Ka Version 07 observations are available through NASA GES DISC. FY-4B AGRI observations are available to the global community on the National Satellite Meteorological Center (NSMC) satellite data server at \url{http://satellite.nsmc.org.cn}. The ground station precipitation data used for validation are provided by the China Meteorological Administration (CMA) through the National Meteorological Information Center (NMIC). Restrictions apply to the availability of these data, which were used under license for the current study and are not publicly available. Access can be obtained with the approval of the CMA via a formal request.
\item Code availability: The source code implementing the PRISMA framework, including the tokenizer architectures, the rectified flow backbone, the conditional branches, and the multi-source posterior sampling pipeline, together with the model configuration files and training/inference scripts used to reproduce the results reported in this study, will be released in a public repository upon acceptance of this manuscript. Pretrained model weights for the AGRI, GMI, SSMIS, and DPR-Ka tokenizers and conditional branches will be deposited in a persistent archive with a DOI for long-term access. Until release, the code and weights are available from the corresponding authors upon reasonable request for the purpose of peer review.
\end{itemize}

\clearpage


\bibliography{sn-bibliography}% common bib file

\end{document}